\documentclass[aps,prl,reprint,twocolumn,superscriptaddress,preprintnumbers,nofootinbib,amsmath,amssymb,tighten]{revtex4-1} 
\usepackage{epsfig}
\usepackage{amsfonts}
\usepackage{amsmath}
\usepackage{amssymb}
\usepackage{dsfont}
\usepackage{hyperref}
\usepackage{xspace}
\usepackage{slashed}
\hypersetup{
    colorlinks=true,       
    linkcolor=blue,          
    citecolor=blue,        
    filecolor=blue,      
    urlcolor=blue           
}
\usepackage{xcolor}
\usepackage{color}
\usepackage{graphicx}  %
\usepackage{bm}  %

\usepackage{graphics}

\renewcommand{\vec}[1]{{\mathbf #1}} 

\newcommand{\simge}{\hspace*{0.2em}\raisebox{0.5ex}{$>$}
     \hspace{-0.8em}\raisebox{-0.3em}{$\sim$}\hspace*{0.2em}}
\newcommand{\simle}{\hspace*{0.2em}\raisebox{0.5ex}{$<$}
     \hspace{-0.8em}\raisebox{-0.3em}{$\sim$}\hspace*{0.2em}}

\newcommand{\sq}{^{2}}

\newcommand{\MS}{$\overline{\rm MS}$ }

\newcommand{\dslash}[1]{#1 \llap{/\kern-0.5pt}}
\newcommand{\Dslash}[1]{#1 \llap{/\kern+1.5pt}}
\newcommand{\DDslash}[1]{#1 \llap{/\kern+2.3pt}}
\newcommand{\dslashh}[1]{#1 \llap{/\kern+1pt}}

\newcommand{\boldtau}{\mbox{\boldmath $\tau$}}
\newcommand{\boldpi}{\mbox{\boldmath $\pi$}}

\newcommand{\boldsigma}{\mbox{\boldmath $\sigma$}}

\newcommand{\bea}{\begin{eqnarray}}
\newcommand{\eea}{\end{eqnarray}}
\newcommand{\be}{\begin{equation}}
\newcommand{\ee}{\end{equation}}
\newcommand{\bma}{\begin{pmatrix}}
\newcommand{\ema}{\end{pmatrix}}

\newcommand{\NLDBD}{$0 \nu \beta \beta$}

\begin{document}
\preprint{LA-UR-18-21404}
\preprint{NIKHEF 2018-010}

\title{
\vspace*{0.5cm}
A new leading contribution to neutrinoless double-beta decay
}

\author{Vincenzo Cirigliano}
\affiliation{Theoretical Division, Los Alamos National Laboratory, Los Alamos, NM 87545, USA}

\author{Wouter Dekens}
\affiliation{Theoretical Division, Los Alamos National Laboratory, Los Alamos, NM 87545, USA}

\author{Jordy de Vries}
\affiliation{Nikhef, Theory Group, Science Park 105, 1098 XG, Amsterdam, The Netherlands}

\author{Michael  L.  Graesser}
\affiliation{Theoretical Division, Los Alamos National Laboratory, Los Alamos, NM 87545, USA}

\author{Emanuele Mereghetti}
\affiliation{Theoretical Division, Los Alamos National Laboratory, Los Alamos, NM 87545, USA}

\author{Saori Pastore}
\affiliation{Theoretical Division, Los Alamos National Laboratory, Los Alamos, NM 87545, USA}

\author{Ubirajara van Kolck}
\affiliation{Institut de Physique Nucl\'eaire, CNRS/IN2P3, Universit\'e Paris-Sud, Universit\'e Paris-Saclay, 91406 Orsay, France}
\affiliation{Department of Physics, University of Arizona, Tucson, Arizona 85721, USA}

\begin{abstract}

Within the framework of chiral effective field theory we discuss the leading 
contributions to the neutrinoless double-beta decay 
transition operator induced by light Majorana neutrinos.  
Based on renormalization arguments in both dimensional regularization 
with minimal subtraction and a coordinate-space cutoff scheme,   
we show the need to introduce a leading-order short-range operator,
missing in all current calculations.    
We discuss strategies to determine the finite part of the short-range 
coupling by matching to lattice QCD or by relating it via chiral symmetry to 
isospin-breaking observables in the two-nucleon sector. 
Finally, we speculate on the impact of this new contribution on nuclear matrix 
elements of relevance to experiment. 

\end{abstract}
\maketitle

\textbf{Introduction:}  Neutrinoless double-beta decay ($0\nu\beta\beta$)  
is the most sensitive laboratory probe of lepton number violation (LNV).
In \NLDBD\  $L$ is violated by two units when
two neutrons in a nucleus turn into two protons, with the emission 
of two electrons and no neutrinos.
The observation of $0\nu\beta\beta$  would 
demonstrate that neutrinos are Majorana fermions~\cite{Schechter:1981bd}, 
shed light on the mechanism of neutrino mass 
generation~\cite{Minkowski:1977sc,Mohapatra:1979ia,GellMann:1980vs}, 
and give insight into leptogenesis scenarios for the generation of the 
matter-antimatter asymmetry in the universe~\cite{Davidson:2008bu}. 

\NLDBD\ is actively being searched for in a number of even-even nuclei 
for which   single-$\beta$ decay is energetically forbidden. 
Current experimental limits~\cite{Gando:2012zm,Agostini:2013mzu,Albert:2014awa,Andringa:2015tza,KamLAND-Zen:2016pfg,Elliott:2016ble,Agostini:2017iyd,Albert:2017owj,Alduino:2017ehq,Azzolini:2018dyb} 
on the half-lives are at the level of $T_{1/2} > 5.3\times10^{25}$~y 
for $^{76}$Ge~\cite{Agostini:2017iyd}
and $T_{1/2} > 1.07\times10^{26}$~y  for $^{136}$Xe~\cite{KamLAND-Zen:2016pfg}, 
with next-generation ton-scale experiments aiming at 
improvements in sensitivity by two orders of magnitude. 

\NLDBD\ can be generated by a variety of dynamical LNV mechanisms, 
which in an effective field theory (EFT) approach to 
new physics are parametrized by $\Delta L=2$ operators of odd dimension
greater than four~\cite{Weinberg:1979sa,Babu:2001ex,Prezeau:2003xn,deGouvea:2007qla,Lehman:2014jma,Graesser:2016bpz,Cirigliano:2017djv}.
If the mass scale associated with LNV is much higher than the electroweak 
scale, the only low-energy manifestation of this new physics is a Majorana mass 
for light neutrinos, encoded in a single gauge-invariant dimension-five 
operator~\cite{Weinberg:1979sa}, which induces \NLDBD\ through light 
Majorana-neutrino exchange~\cite{Bilenky:1987ty,Bilenky:2014uka}.  
To interpret positive or null \NLDBD\ results in this minimal scenario it is  
crucial to have good control over the relevant hadronic and nuclear matrix 
elements.  Current knowledge of these is not satisfactory~\cite{Engel:2016xgb},
as   various many-body approaches lead to estimates that differ by a factor 
of two to three and most calculations are not based on a modern 
EFT analysis.  
In Ref.~\cite{Cirigliano:2017tvr} a first step was presented towards the 
analysis of \NLDBD\ induced by a light Majorana neutrino in the chiral 
EFT framework~\cite{Bedaque:2002mn,Epelbaum:2008ga,Machleidt:2011zz}, which 
provides a systematic expansion of hadronic amplitudes in 
$p / \Lambda_\chi$, where $p \sim m_\pi \sim k_F \sim {\cal O}(100~{\rm MeV})$ 
and 
$\Lambda_\chi  \sim 4\pi F_\pi  \sim m_N \sim {{\cal O}(1~{\rm GeV})}$. 
The \NLDBD\ transition operators were derived 
up to next-to-next-to-leading order (N$^2$LO) in 
Weinberg's power-counting scheme~\cite{Weinberg:1990rz,Weinberg:1991um}.

In this letter we demonstrate that Weinberg's scheme for \NLDBD\  
assumed in Ref.~\cite{Cirigliano:2017tvr} 
breaks down and any consistent power counting requires a leading-order (LO) 
short-range $\Delta L=2$ operator, 
whose effect is missing in all current calculations.  
Our argument is based on renormalization. 
Using two different schemes (dimensional regularization with minimal 
subtraction and a coordinate-space cutoff) 
we show that  once the strong nucleon-nucleon scattering amplitude is made 
finite and independent of the ultraviolet regulator, 
an additional $\Delta L=2$ contact operator with coupling $g_\nu^{N\!N}$ 
has to be introduced to make the $nn \to p p ee$ amplitude finite and 
regulator-independent.   
The finite part of $g_\nu^{N\!N}$, which encodes hard-neutrino exchange, 
can be determined by (i) matching the chiral EFT
$nn \to p p ee$ amplitude to future lattice QCD calculations; 
(ii) relating it via chiral symmetry
to electromagnetic low-energy constants (LECs) that control isospin-breaking 
in the two-nucleon sector.
A combination of couplings involving $g_\nu^{N\!N}$ can be fit to nucleon-nucleon
charge-independence-breaking (CIB) observables, 
confirming the LO scaling of this coupling.   
Based on this, we speculate on the impact of $g_\nu^{N\!N}$ on nuclear matrix 
elements of relevance to experiments.

\textbf{The need for an LO short-range $\Delta L =2$ interaction:}
We consider a scenario in which LNV at low energy is dominated by  
the electron-neutrino Majorana mass
\begin{equation}
\mathcal L_{\Delta L =2} = - \frac{m_{\beta\beta}}{2} \, \nu^T_{e L} C \nu_{eL},
\end{equation}
where $C= i \gamma_2 \gamma_0$ denotes the charge conjugation matrix. 

The nuclear effective Hamiltonian can be written as 
\be
H_{\rm eff} =  H_{\rm strong} +  2 G_F^2  V_{ud}^2 \  m_{\beta \beta}  
\  \bar e_{L} C \bar e_{L}^T   \ V_\nu  \,,
\label{eq:HV}
\ee
in terms of the Fermi constant $G_F$
and  the $V_{ud}$ element of the  CKM 
matrix~\cite{Cabibbo:1963yz,Kobayashi:1973fv}. 
The neutrino potential $V_\nu$
can be obtained from
two-nucleon irreducible diagrams mediating $nn \to pp ee$ 
to a given order in $p/\Lambda_\chi$. 
Within Weinberg's power counting the only LO 
contribution~\cite{Cirigliano:2017tvr}
comes from the exchange of potential neutrinos, 
with $q^0  \ll |\vec q|$,
\begin{eqnarray}\label{nupot}
V_{\nu,0}(\vec q) &=& \tau^{(1) +} \tau^{(2) +} \frac{1}{\vec q^2}  \Bigg\{ 1 - g_A^2 \boldsigma^{(1)} \cdot \boldsigma^{(2)}
\nonumber \\
&+&  
g_A^2 \, \boldsigma^{(1)} \cdot \vec q\, \boldsigma^{(2)}\cdot \vec q  \    
\frac{2 m_\pi^2  + \vec{q}^2}{(\vec q^2 + m_\pi^2)^2} \Bigg\}~, 
\end{eqnarray}
where $g_A \simeq 1.27$ is the nucleon axial coupling, $m_\pi$ the pion mass, 
and $\vec q$ the momentum transfer. 
N$^2$LO terms arise from corrections to the single nucleon weak currents,  
irreducible one-loop diagrams, and contact interactions   
mediating $\pi \pi \to ee$, $n \to p \pi^+  e e$, and $nn \to ppee$.
In particular, the short-range potential includes a 
two-nucleon term~\cite{Cirigliano:2017tvr}
\begin{equation}\label{eq:VCT}
V_{\nu, CT} =  -  2 g_\nu^{N\!N}  \  \tau^{(1) +} \tau^{(2) +}   \,,
\end{equation}
where the LEC $g_{\nu}^{N\!N}$ is 
$\mathcal O((4 \pi F_\pi)^{-2})$   
in Weinberg's counting and $F_\pi = 92.2$~MeV is the pion decay constant.
However, it is known that 
Weinberg's power counting leads to inconsistent results in 
nucleon-nucleon  scattering~\cite{Kaplan:1996xu,Beane:2001bc,Nogga:2005hy,Long:2012ve}  
and nuclear processes mediated by external currents~\cite{Valderrama:2014vra},
due to a conflict between  naive dimensional analysis and
nonperturbative renormalization. 
We therefore investigate the scaling of $g_\nu^{N\!N}$
by studying the amplitude 
$\mathcal A (n n \rightarrow p p ee) \equiv \mathcal A_{\Delta L=2}$ 
with strong interactions, $H_{\rm strong}$, included nonperturbatively.

We work at LO in chiral EFT, and focus on the scattering of two neutrons to 
two protons in the $^1S_0$  wave, where $H_{\rm strong}$ has 
short-range and Yukawa  components, 
\begin{equation}\label{eq:1}
V_0(\vec q) = \tilde{C} + V_{\pi}(\vec q) \, ,
\quad  
V_{\pi}(\vec q) = -
\frac{g_A^2}{4 F_\pi^2}\frac{m_\pi^2}{\vec q^2 + m_\pi^2}\,,
\end{equation}
with  $\tilde C \sim \mathcal O(F_\pi^{-2}, m_\pi^2F_\pi^{-4})$~\cite{Weinberg:1991um,Kaplan:1996xu,Beane:2001bc}.
We have checked that transitions involving higher partial waves
such as ${}^3P_{0,1}\rightarrow {}^3P_{0,1}$ are correctly renormalized and do 
not require enhanced $\Delta L=2$ counterterms.

\begin{figure}
\includegraphics[width=0.5\textwidth]{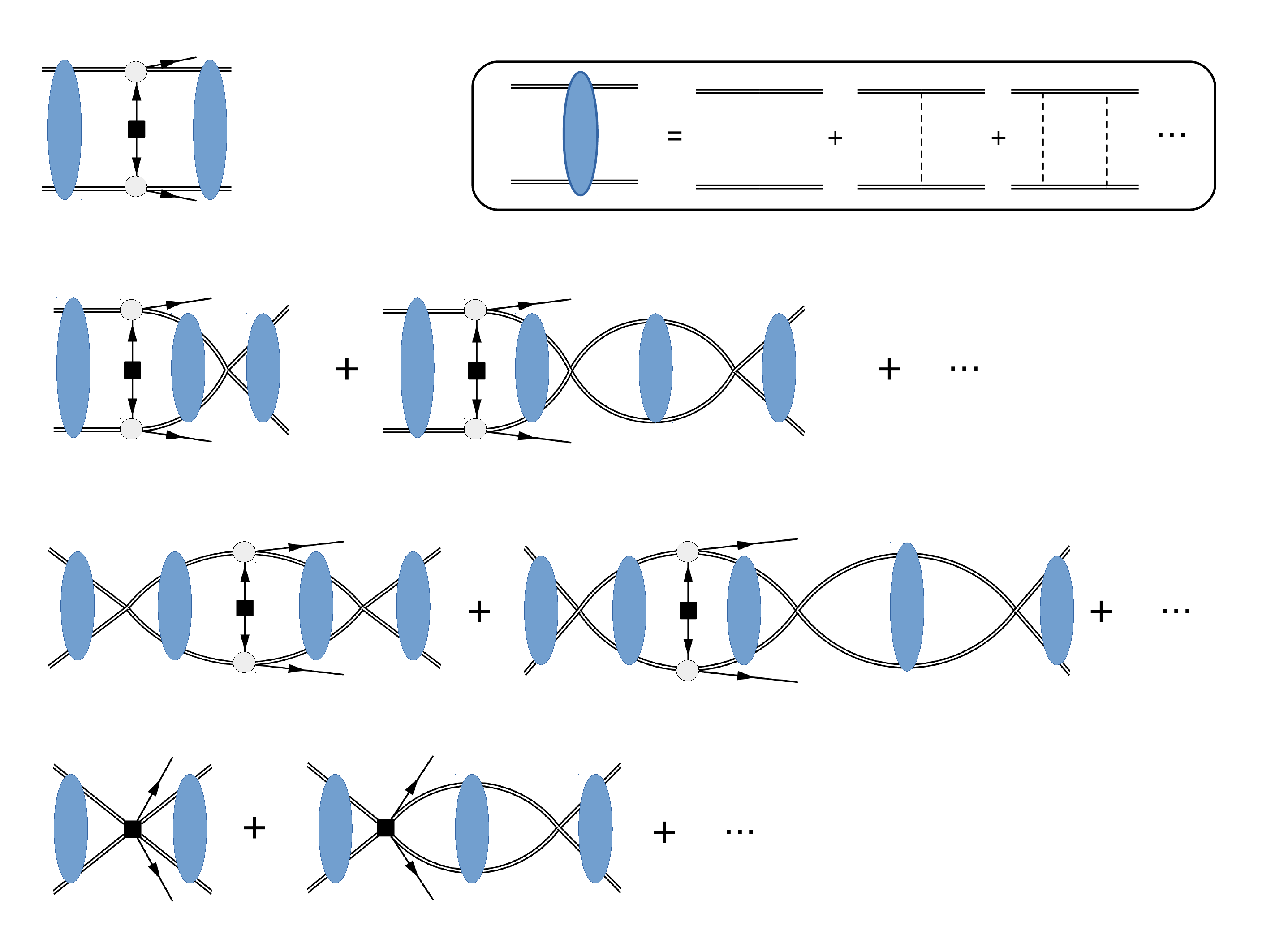}
\caption{Diagrammatic representation of LO contributions to 
$n n \rightarrow p p ee$. Double, dashed, and plain lines denote nucleons, 
pions, and leptons, respectively. 
Gray circles denote the nucleon axial and vector currents, and 
the black square an insertion of $m_{\beta \beta}$. 
The blue ellipse represents iteration of 
$V_\pi$. In the counterterm amplitude (fourth line) the black square represents
$g^{N\!N}_\nu$. The $\cdots$ in the second to fourth lines denote diagrams with 
arbitrary numbers of bubble insertions.  
}\label{Fig1}
\end{figure}

The contributions to  $\mathcal A_{\Delta L=2}$
from the exchange of a light neutrino ($\mathcal A_{\Delta L=2}^{(\nu)}$)
are shown in Fig.~\ref{Fig1}. 
The blue ellipse denotes the iteration of the Yukawa potential 
$V_\pi(\vec q)$. 
The diagrams in the second and third rows
include an infinite number of bubbles, dressed with 
iterations of $V_\pi$.
Without loss of generality for our arguments, we use the kinematics 
$n(\vec p) \ n (-\vec p) \to p (\vec p^\prime) \ p (- \vec p^\prime) 
\ e ( \vec{p_{e1}} =0) \ e ( \vec{p_{e2}} =0) $, 
with $|\vec{p} |= 1$~MeV and correspondingly $ |\vec{p} '| = 38$~MeV.

$\mathcal A_{\Delta L=2}^{(\nu)}$ can be expressed  in terms of the Yukawa ``in'' 
and ``out'' wavefunctions $\chi_{\vec p}^\pm (\vec r)$ and the propagators 
$G^\pm_{E}(\vec r, \vec r^\prime) = \langle \vec r^\prime | 
(E - T - V_\pi \pm i  0^+)^{-1}| \vec r \rangle$~\cite{Kaplan:1996xu,Long:2012ve}.
Observing that the bubble diagrams in Fig.~\ref{Fig1} are related to 
$G^+_E(\vec 0,\vec 0)$, while the triangles dressed by Yukawas are related to 
$\chi^+_{\vec p}(\vec 0)$ and 
$\chi^-_{\vec p^\prime}(\vec 0)^* =\chi^+_{\vec p^\prime}(\vec 0)$~\cite{Kaplan:1996xu},
the LO amplitude reads
\begin{eqnarray}
\mathcal A_{\Delta L=2}^{(\nu)}  &=& \mathcal A_A  
+ K_{E'} \, \mathcal A_B + \mathcal {\bar A}_B \, K_{E}  
+ K_{E'} \, \mathcal A_C \, K_{E}   \,,
\nonumber \\
K_{E} &=& \frac{    \chi^+_{\vec p}(\vec 0) \ 
\tilde{C}}{1 - \tilde C G^+_{E}(\vec 0, \vec 0)}  \,,
\label{eq:amplitude1}
\end{eqnarray}
where $\mathcal A_A$, $\mathcal A_B$, and $\mathcal A_C$ denote the first 
diagram in the first, second, and third rows of Fig.~\ref{Fig1}, 
respectively (without the wavefunctions at $\vec{0}$, in the case of $\mathcal A_B$ and $\mathcal A_C$).  
$\mathcal {\bar A}_B$ is similar to $\mathcal A_B$ and not shown in Fig. \ref{Fig1}.

To study the renormalization of the $\Delta L =2$ amplitude,  we now discuss the divergence structure of 
$ \mathcal A_{\Delta L=2}^{(\nu)}$. 
$ \chi^+_{\vec p}(\vec 0)$ is finite and the divergence in 
$G^+_{E}(\vec 0, \vec 0)$ is absorbed by $\tilde{C}^{-1}$, so that  
$K_E$ is finite and scheme-independent~\cite{Kaplan:1996xu}.
We note that:

(i) All diagrams in $\mathcal A_A$ are finite. The tree level is  finite and 
each $V_\pi$ iteration 
improves the convergence by bringing in a factor of $d^3 \vec k/(\vec k^2)^2$,
where one $\vec k^2$ comes from the pion propagator and the other from the 
two-nucleon propagator.

(ii) All the diagrams in $\mathcal A_B$ and $\bar{\mathcal A}_B$ are finite. 
The first loop goes as $d^3 \vec k/(\vec k^2)^2$, while 
$V_\pi$ insertions 
further improve the convergence.

(iii) The first  two-loop  diagram in $\mathcal A_C$  has a logarithmic 
divergence, which stems from an insertion of the most singular component 
of the neutrino potential, namely 
\be
\tilde{V}_{\nu}(\vec q) = \tau^{(1) +} \tau^{(2) +} \frac{1}{\vec q^2} 
\left( 1 - \frac{2}{3} g_A^2 \boldsigma^{(1)} \cdot \boldsigma^{(2)} \right)~.
\ee
The two-loop diagram with insertion of $V_{\nu,0} - \tilde{V}_\nu$ and 
higher-loop diagrams are convergent. 

We focus on $\mathcal A_C$ and write
$\mathcal A_C =  \mathcal{A}_C^{(\rm div)} +  \delta \mathcal A_C$. 
In dimensional regularization,
\bea
\mathcal{A}_C^{\rm (div)}  &=& -\left(\frac{m_N}{4\pi}\right)^2 
\left(1 + 2 g_A^2\right)
\Big[  \Delta  + L_{\vec{p},\vec{p}'} (\mu)  \Big] \,,
\\
L_{\vec{p},\vec{p}'} (\mu) &=& \frac{1}{2} 
\left(\log \frac{\mu^2}{ -(  |\vec{p}| + |\vec{p}'|)^2  + i 0^+ }  + 1 \right)~, 
\nonumber
\eea
where $\Delta \equiv  \left(  1/(4 -d)   - \gamma + \log 4 \pi \right) /2$. 
The divergence for $d \to 4$ can be removed by introducing 
$g_\nu^{N\!N}$ at LO. 
The counterterm amplitude, shown in the fourth line of Fig.~\ref{Fig1}, reads
\begin{equation}\label{eq:Ann}
\mathcal A_{\Delta L=2}^{(N\!N)} = K_{E'} \ \frac{2 g_\nu^{N\!N}}{\tilde{C}^2} \ K_E \,,
\end{equation}
and we can renormalize $\mathcal{A}_{\Delta L=2}$ by replacing  
$\mathcal A_C \to \mathcal A_C +  2 g_\nu^{N\!N}/\tilde C^2$
in Eq.~\eqref{eq:amplitude1}.    
In the $\overline{\rm MS}$ scheme,
\be
\mathcal A_C \to \delta \mathcal A_C 
+ \left( \frac{m_N }{4\pi} \right)^2
\Big[ 2 \tilde{g}_{\nu}^{N\!N}(\mu) 
- \left( 1 + 2 g_A^2 \right) L_{\vec{p},\vec{p}'} (\mu) \Big]  
\label{eq:amplitude2}
\ee
after defining the dimensionless coupling 
\be
\tilde{g}_\nu^{N\!N} = 
\left(\frac{4\pi}{m_N\tilde C}\right)^2 g_\nu^{N\!N} \,.
\label{eq:gnutilde}
\ee
This coupling obeys the renormalization-group equation (RGE)
\be
\mu \frac{d\tilde{g}_\nu^{N\!N}}{d\mu} = \frac{1}{2} \left(1 + 2 g_A^2\right)~,
\ee
confirming that $\tilde{g}_\nu^{N\!N} \sim \mathcal O(1)$.
Since $\tilde C(\mu=m_\pi)\approx -0.9/F_\pi\sq$,
we find that $g_\nu^{N\!N} \sim \mathcal O\left( F_\pi^{-2} \right)$ instead of
$\mathcal O\left( (4 \pi F_\pi)^{-2} \right)$.
A similar enhancement also occurs in 
four-nucleon couplings induced by higher-dimensional LNV operators.
Treating $V_\pi$ as a subleading correction~\cite{Kaplan:1998we,Beane:2001bc} 
is equivalent to working to LO in pionless EFT,  and does not affect our conclusions about the importance of $g_\nu^{N\!N}$~\cite{Cirigliano:2017tvr}.
Details on how to obtain  $\delta \mathcal A_C$ will be provided in future work~\cite{future}.

\textbf{$\mathcal A_{\Delta L=2}$ in a cutoff scheme:}
The need for an LO  counterterm can be demonstrated also in a coordinate-space 
scheme that makes no  direct
reference to Feynman diagrams. In this approach we regulate  
the short-range part of $V_0$
with a smeared $\delta$-function,
\begin{equation}
\tilde C \, \delta^{(3)}(\vec r) \rightarrow  
\frac{\tilde{C}  (R_S)}{(\sqrt{\pi}R_S)^3} 
\exp{\left(-\frac{r^2}{R_S^2}\right)} 
\equiv \tilde C  {(R_S)} \ \delta_{R_S}^{(3)} (\vec r)~, 
\label{Gaussreg}
\end{equation}
and  obtain $\psi^-_{\vec p^\prime} (\vec r)$ and $\psi^+_{\vec p} (\vec r)$
by solving the Schr\"odinger equation.
We  determine $\tilde C (R_S)$ by requiring that the $^1S_0$ scattering length 
be reproduced  ($\tilde C \approx -0.4/F_\pi\sq$ at $R_S=0.8$ fm).
We find that $1/\tilde C (R_S)$ has  linear ($1/R_S$) and 
logarithmic divergences \cite{Beane:2001bc}  
and that the $^1S_0$ phase shifts at nonzero momentum are indeed 
$R_S$-independent.

\begin{figure}
\includegraphics[width=0.45\textwidth]{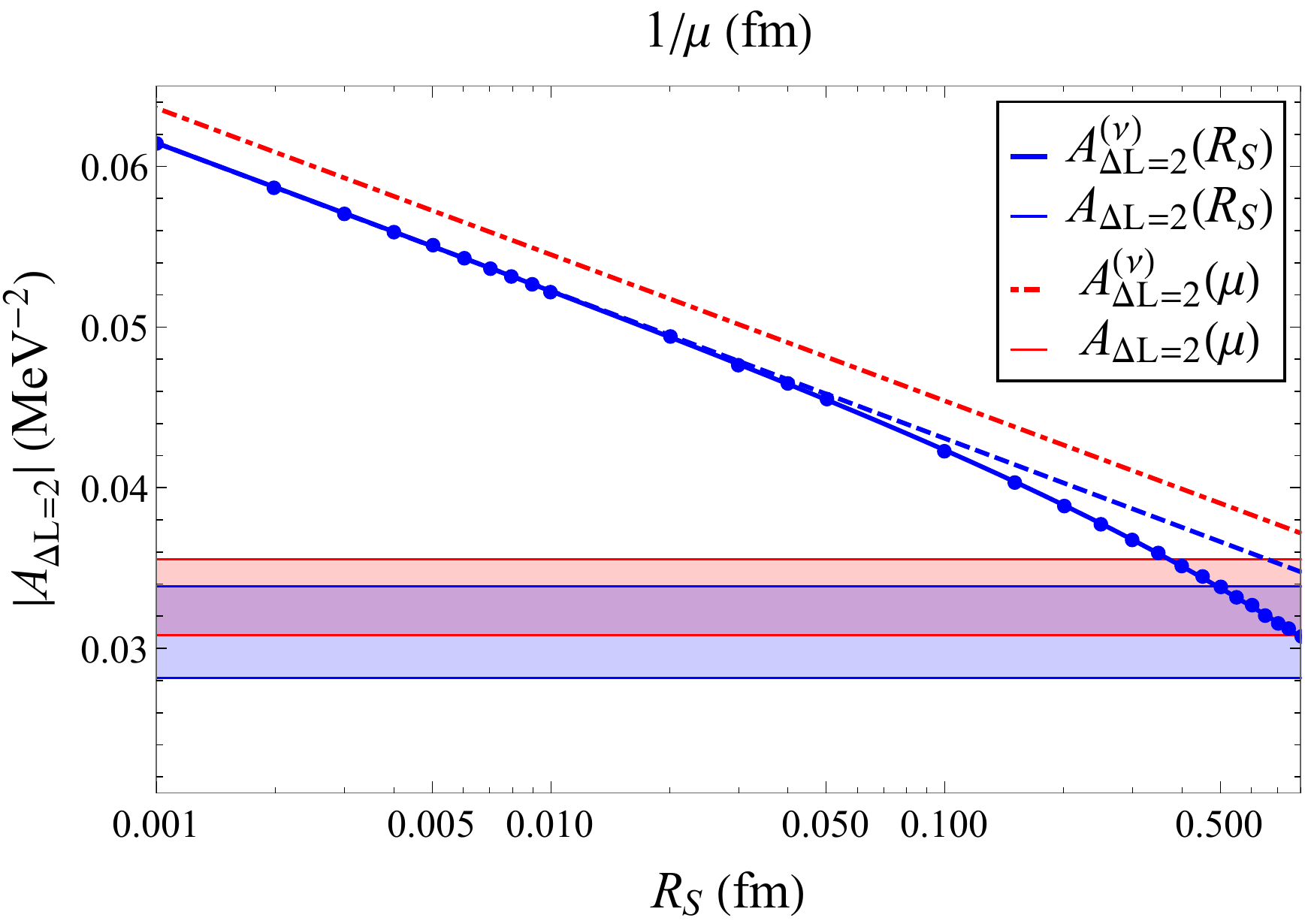}
\caption{Matrix element $\mathcal A_{\Delta L=2}^{(\nu)}$ for $|\vec p| =1$ MeV and
$|\vec p^\prime|= 38$ MeV, as a function of $R_S$. 
The dashed line shows a fit to $a + b \log R_S$, which captures the small 
$R_S$ behavior.
The solid line corresponds to a 
fit that includes ${\cal O}(R_S, R_S \log R_S)$ power corrections. 
The dash-dotted line shows $\mathcal A_{\Delta L=2}^{(\nu)}$ in 
\MS as a function of $1/\mu$.
The horizontal bands represent the total amplitude $\mathcal A_{\Delta L=2}$   
with $g_{\nu}^{N\!N} =(C_1+C_2)/2$, as discussed in the main text. 
}
\label{Fig3}
\end{figure}

We then compute 
\begin{eqnarray}\label{eq:anutot_v1}
\mathcal A_{\Delta L=2}^{(\nu)} = - \int\! d^3\vec r \ \psi^-_{\vec p^\prime}(\vec r)^* 
\  V_{\nu,0}(\vec r)  \ \psi^+_{\vec p}(\vec r)~,
\end{eqnarray}
where $V_{\nu,0} (\vec r)$ is obtained by Fourier-transforming the 
$^1 S_0$ projection of Eq.~\eqref{nupot}. 
In Fig.~\ref{Fig3} we plot $\mathcal A_{\Delta L=2}^{(\nu)}$ as a function of $R_S$. 
The plot displays a logarithmic dependence on $R_S$ 
(analogous to the $\log \mu$ dependence in Eq.~\eqref{eq:amplitude2}) 
as well as milder power-like behavior. Therefore, to obtain a physical,
regulator-independent amplitude one needs to include an LO counterterm,  given in $r$-space by 
$V_{\nu,CT}(\vec r)= - 2 \,g_\nu^{N\!N} (R_S)\delta_{R_S}^{(3)} (\vec r)$.   
The corresponding amplitude,
\be
\mathcal A_{\Delta L=2}^{(N\!N)} =  - \int \! d^3 \vec r  \    
\psi^-_{\vec p^\prime}(\vec r)^* \ V_{\nu,CT}(\vec r) \ \psi^+_{\vec p}(\vec r)
\, ,
 \label{eq:aNN}
\ee
is also regulator-dependent. 
As expected from Eq. \eqref{eq:Ann}, we find its leading divergent behavior  
to be well reproduced by $1/\tilde C(R_S)^2$. 
We can then make  
$\mathcal A_{\Delta L=2} =\mathcal A_{\Delta L=2}^{(\nu)}+\mathcal A_{\Delta L=2}^{(N\!N)}$ 
finite for $R_S \to 0$ and $R_S$-independent by choosing 
$\tilde{g}_\nu^{N\!N} (R_S) = -(a/2) (1 + 2 g_A^2) \log R_S + b + c R_S +  \cdots$, 
with the coefficient of the logarithm quite close to the 
$\overline{\textrm{MS}}$ expectation $a=1$.

\textbf{Relating $g_\nu^{N\!N}$  to electromagnetic isospin violation:}   
The finite part of $g_\nu^{N\!N}$ can be obtained  by matching 
the chiral EFT amplitude to a lattice QCD calculation performed at the same 
kinematic point, as it is done in the strong-interacting sector~\cite{Barnea:2013uqa}.
First lattice results related to double-beta decay are starting to appear~\cite{Nicholson:2016byl,Shanahan:2017bgi}.

We now discuss a complementary estimate 
based on the fact that the short-range operators and associated LECs arising 
in \NLDBD\ and electromagnetic processes 
are closely related~\cite{Cirigliano:2017tvr}. 
In the electromagnetic case, the short-range hadronic operators arise from   
amplitudes in the underlying theory involving  two insertions of the 
electromagnetic current with  exchange of hard virtual 
photons~\cite{Moussallam:1997xx,Ananthanarayan:2004qk}. 
In the $\Delta L=2$ case, up to a proportionality factor, the {\it same} 
operators are generated by the insertion of two weak currents with 
exchange of hard neutrinos. This comes about because the neutrino propagator 
and weak vertices combine to give a massless gauge-boson propagator 
in Feynman gauge, multiplied by 
$8 G_F^2 V_{ud}^2 m_{\beta \beta} \bar{e}_L e_L^c$~\cite{Cirigliano:2017tvr}.   
The LECs needed for \NLDBD\ are therefore related to the  LECs associated 
with the isospin $I = 2$ component of the product of two 
electromagnetic currents,  
which belongs to the  $5_L\times 1_R$ irreducible representation of 
chiral $SU(2)_L\times SU(2)_R$.

Only two independent four-nucleon operators that transform as 
$I=2$ objects exist: 
\begin{eqnarray}\label{eq:4N}
O_1 &=& \bar N \mathcal Q_L N \, \bar N \mathcal Q_L N  
-\frac{\textrm{Tr}[\mathcal Q^2_L]}{6} \bar N \boldtau N \cdot \bar N \boldtau N
+ \{ L \leftrightarrow R \} \, ,
\nonumber \\
O_2 &=&  2 \left( \bar N \mathcal Q_L N \bar N \mathcal  Q_R N  
-\frac{\textrm{Tr}[\mathcal Q_L \mathcal Q_R] }{6} \bar N \boldtau N \cdot 
\bar N \boldtau N \right) \, ,
\end{eqnarray}
where $\mathcal Q_L  = u^\dagger Q_L u$, 
$\mathcal Q_R  = u Q_R u^\dagger$, 
$u = \exp(i \boldtau \cdot \boldpi/(2 F_{\pi}))$, 
and $Q_{L,R}$ are ``spurions" transforming under 
the chiral group as $Q_L  \rightarrow L Q_L  L^\dagger$,  $Q_R \rightarrow RQ_R R^\dagger$. 
In the electromagnetic case $Q_L = Q_R = \tau_3/2$, 
while in $0\nu\beta\beta$ $Q_L = \tau^+$, $Q_R = 0$. 
In our conventions $O_1$ enters the $\Delta L=2$ Lagrangian  
with coefficient  $2 G_F^2 V_{ud}^2 m_{\beta\beta} g_\nu^{N\!N} $. 
Defining the electromagnetic LECs multiplying $O_{1,2}$ as $e^2 C_{1,2}/4$, 
chiral symmetry dictates $g^{N\!N}_{\nu} = C_1$. 

In the electromagnetic case, 
$O_1$ and $O_2$ only differ at the multipion level, and an isospin-breaking 
two-nucleon observable, such as the $I=2$ combination of scattering 
lengths $a_{\rm CIB} = (a_{nn} + a_{pp})/2 - a_{np}$, only constrains the sum 
$C_1 + C_2$. 
Extracting this combination from data provides 
a rough estimate of $g_\nu^{N\!N}$ under the assumption $C_1 \sim C_2$.  
As in the $\Delta L=2$ case, we introduce the dimensionless couplings 
$\tilde C_i \equiv [4\pi/(m_N \tilde C)]^2 C_i$  
and  compute the scattering lengths $a_{pp,nn,np}$ including the leading 
sources of isospin breaking -- the Coulomb potential and pion mass splitting --
and $\tilde C_1 + \tilde C_2$.  
Similarly to the $\Delta L=2$ case, we find that $\tilde C_1 + \tilde C_2$  
needs to be promoted to LO and obeys the RGE
\begin{eqnarray}\label{DimRegEM}
\mu\frac{d}{d\mu}
\frac{\tilde C_1 + \tilde C_2}{2} &=&  \frac{1}{2}  
\left( 1 + g_A^2\, \frac{m^2_{\pi^+}  - m^2_{\pi^0}}{e^2F_\pi^2}\right)\,,
\end{eqnarray}
while, of course, $\tilde C_1$ has the same RGE as $\tilde{g}_\nu^{N\!N}$.
By fitting to $a_{\rm CIB}$ using  
$a_{np} = -23.7$~fm, 
$a_{nn} = -18.9$~fm,
and $a_{pp} = -7.8$~fm, 
we find $(\tilde C_1+\tilde C_2)/2 \approx 2.5$  
at $\mu = m_\pi$ in the \MS scheme. 
Using instead the $R_S$ scheme, we find $(\tilde C_1+\tilde C_2)/2 \approx 2.0$
at $R_S = 0.5$~fm.~\footnote{Our result is consistent with analyses based on 
chiral~\cite{Walzl:2000cx, Piarulli:2014bda,Piarulli:2016vel,Reinert:2017usi} 
and phenomenological potentials such as AV18~\cite{Wiringa:1994wb}, 
which also find that, except at very low energies, 
long- and short-range components of the CIB interaction 
induce effects of similar size.} 
This estimate, based on data and chiral symmetry, again confirms that 
$g_\nu^{N\!N} \sim \mathcal O(F_\pi^{-2})$.

\begin{figure}[t]
\includegraphics[width=0.4\textwidth , trim={0 .2cm  2.1cm 2.1cm},clip]{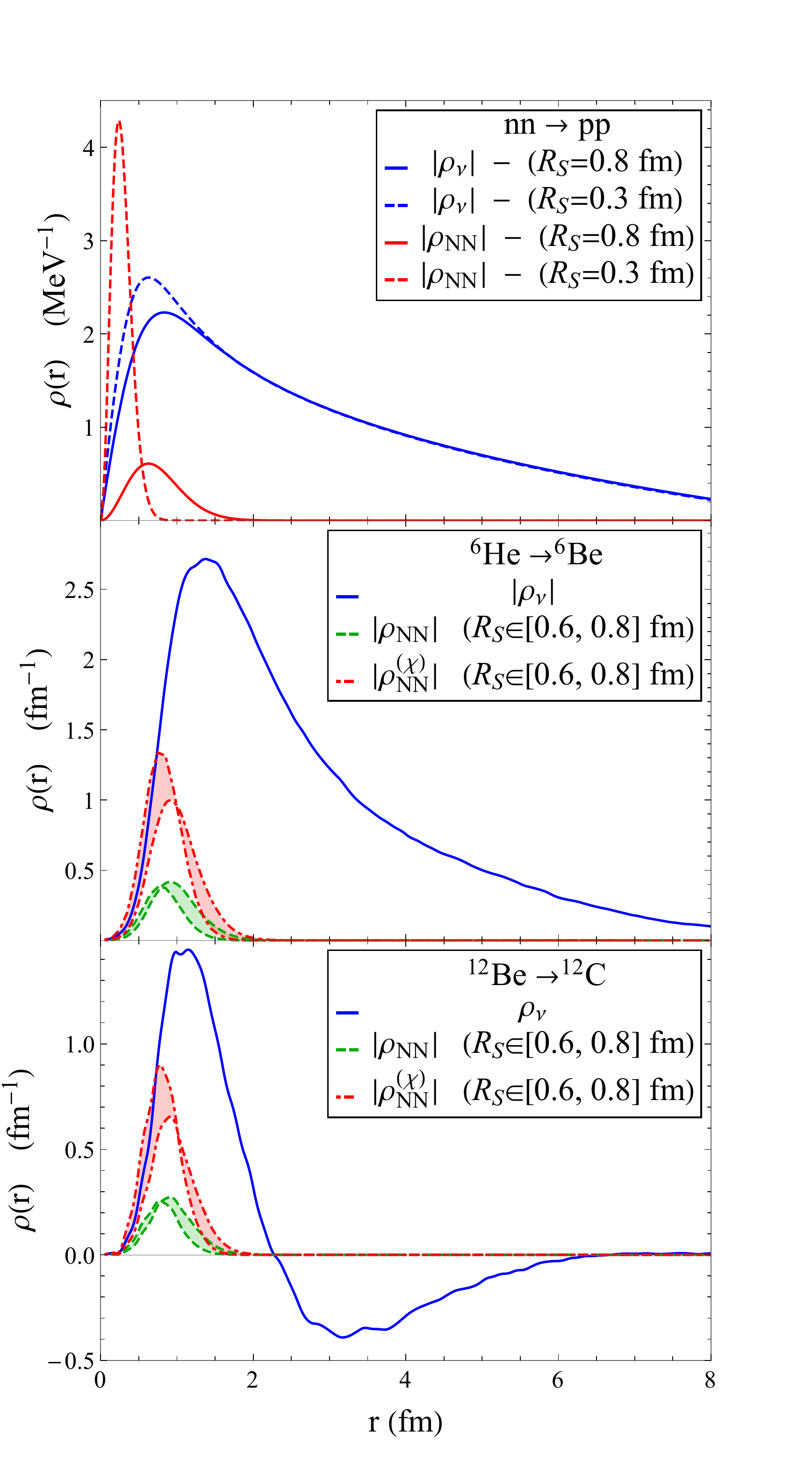}
\hspace{0.03\linewidth}
\caption{$\rho_{\nu}(r)$ and $\rho_{N\!N}(r)$ for the $nn \to ppee$ process (top),
and for nuclear transitions with $A=6$ (middle) and $A=12$ (bottom). 
In the middle and bottom panels the green ($\rho_{N\!N}$)
and red ($\rho_{N\!N}^{(\chi)}$) bands correspond to 
$g_\nu^{NN} = (C_1+C_2)/2$ extracted from our analysis and 
from Refs.~\cite{Piarulli:2014bda,Piarulli:2016vel}, respectively.
}
\label{densityzoom}
\end{figure}

\textbf{Numerical impact:} To roughly  estimate the impact of the contact term, we assume 
for concreteness $C_1 = C_2$ and hence $g_{\nu}^{N\!N}= (C_1 + C_2)/2$ at some   $\bar{R}_S$ or $\bar\mu^{-1}$ 
in the range $0.002 - 0.8$~fm. The total two-nucleon amplitude 
$\mathcal A_{\Delta L=2}=\mathcal A_{\Delta L=2}^{(\nu)} +\mathcal A_{\Delta L=2}^{(N\!N)}$ 
then becomes independent of the regulator, as illustrated in Fig.~\ref{Fig3}, 
where the widths of the horizontal bands reflect the ambiguity in the choice 
of the  point $\bar{R}_S$  or $\bar\mu$ {\it where} $C_1= C_2$ is assumed.
(They do not account for the uncontrolled error of the assumption itself.)
The relative size of the two components depends on $R_S$,  with 
$\mathcal A_{\Delta L=2}^{(N\!N)}/\mathcal A_{\Delta L=2}^{(\nu)}\sim 30\%$ 
at $R_S \sim 0.1$~fm,  
decreasing to $\sim 10\%$ at $R_S \sim 0.6$~fm. 
More insight can be obtained by plotting the matrix-element densities 
$\rho_{\nu}$ and $\rho_{N\!N}$ defined as
\begin{equation}
\mathcal A_{\Delta L=2}^{(\nu)} = \int \! d r \, \rho_\nu(r) \,, 
\quad  
\mathcal A_{\Delta L=2}^{(N\!N)} = \int \! d r \, \rho_{N\!N}(r) \,.
\end{equation}
Figure~\ref{densityzoom} (top panel) shows that 
$\rho_{N\!N} (r)$ is concentrated at smaller  distances 
than $\rho_\nu(r)$, and  its contribution to the amplitude is thus partially  diluted.

We have performed a similar analysis for $A=6, 12$ nuclei,
using Variational Monte Carlo nuclear wavefunctions~\cite{Pastore:2017ofx} 
based on the AV18 two-nucleon~\cite{Wiringa:1994wb} and 
IL7 three-nucleon~\cite{IL7} interactions.  
The mismatch between the short-range behaviors of existing 
strong-interaction potentials and our \NLDBD\ interaction introduces additional
model dependence, which we mitigate by: 
(i) Considering an  alternative extraction of $(C_1+C_2)/2$ from 
the phase-shift analysis of  Refs.~\cite{Piarulli:2014bda,Piarulli:2016vel}~\footnote{ 
$C_1 + C_2$ is related to the CIB coefficient $C_0^{IT}$ of Refs.~\cite{Piarulli:2014bda,Piarulli:2016vel}
by $(C_1+C_2)/2 = - 6 C_0^{IT}/e^2$.},  
which  employs the same regulator \eqref{Gaussreg} with 
$R_S \simeq 0.6 - 0.8$~fm, approximately the range of AV18's short-range  part. 
(ii) Simply replacing our $V_{\nu,CT}(\vec r)$ with AV18's short-range CIB potential.

For $\Delta I = 0$ transitions such as the $^6$He $\rightarrow$ $^6$Be 
shown in Fig.~\ref{densityzoom} (middle panel), we find 
$\mathcal A_{\Delta L=2}^{(N\!N)}/\mathcal A_{\Delta L=2}^{(\nu)}\sim 10\%$,  
similarly to the $n n \rightarrow p p ee$ case.
In realistic $0\nu\beta\beta$ transitions, however, 
the total nuclear isospin changes by two units, $\Delta I = 2$. 
This implies the presence of a node in 
$\rho_\nu(r)$ due to the orthogonality 
of the initial and final spatial  wavefunctions.
The resulting partial cancellation
between the regions with $r \simle  2$~fm and 
$r \simge  2$~fm~\cite{Pastore:2017ofx}  
leads to a relative enhancement of the short-range contribution,
as illustrated in Fig.~\ref{densityzoom} (bottom panel)  
for $^{12}$Be $\rightarrow$ $^{12}$C. 
Numerically we find $\mathcal A_{\Delta L=2}^{(N\!N)}/\mathcal A_{\Delta L=2}^{(\nu)}\sim 25\%$
(our fit),    $\sim 55\%$  (fit from Refs.~\cite{Piarulli:2014bda,Piarulli:2016vel}), 
and  $\sim 60\%$    (AV18 representation of the short-range CIB potential).
Because the node in the density is a robust feature of $\Delta I=2$ 
transition~\cite{Simkovic:2007vu,Menendez:2008jp},  
we expect the effects in  $^{12}$Be $\rightarrow$ $^{12}$C and experimentally 
relevant transitions to be of comparable size.

\textbf{Conclusion:}  The above arguments suggest that the new short-range $\Delta L=2$ 
potential identified in this work can significantly impact \NLDBD\ 
phenomenology   and its implications for Majorana neutrino masses.
We hope this will stimulate work towards a more controlled determination of  $g_\nu^{N\!N}$ from lattice QCD 
and an assessment of the impact of the short-range potential  in nuclei of experimental interest.

\begin{acknowledgments}
\textbf{Acknowledgments:}
VC, WD, MG, EM, and SP acknowledge support by the U.S. Department of Energy, 
Office of Science, and by the LDRD program at Los Alamos National Laboratory.
VC and EM acknowledge partial support from the  DOE topical collaboration on 
``Nuclear Theory for Double-Beta Decay and Fundamental Symmetries''. 
JdV acknowledges support by the Dutch Organization for Scientific Research 
(NWO)  through a VENI grant.  
The work of UvK was supported in part by the U.S. Department of Energy, 
Office of Science, Office of Nuclear Physics, under award No. 
DE-FG02-04ER41338, and by the European Union Research and Innovation program 
Horizon 2020 under grant agreement No. 654002.
We acknowledge  stimulating discussions with Will Detmold and Martin Savage, 
which  triggered this research. We thank Joe Carlson, 
Jon Engel, Amy Nicholson, Maria Piarulli, Petr Vogel,  Andre Walker-Loud, 
and Bob Wiringa for discussions at various stages of this work.  
\end{acknowledgments}

\bibliography{bibliography}

\end{document}